# Geo-knowledge-guided GPT models improve the extraction of location descriptions from disaster-related social media messages [1]

Yingjie Hu [a†*], Gengchen Mai [b†], Chris Cundy [c], Kristy Choi [c], Ni Lao [d], Wei Liu [a], Gaurish Lakhanpal [a,e], Ryan Zhenqi Zhou [a], and Kenneth Joseph [f]

[a] *GeoAI Lab, Department of Geography, University at Buffalo, Buffalo, NY, USA*
[b] *Spatially Explicit Artificial Intelligence (SEAI) Lab, Department of Geography, University of Georgia, Athens, GA, USA*
[c] *Department of Computer Science, Stanford University, Stanford, CA, USA*
[d] *Google, Mountain View, CA, USA*
[e] *Stevenson High School, Lincolnshire, IL, USA*
[f] *Department of Computer Science and Engineering, University at Buffalo, Buffalo, NY, USA*

[†] *These authors contributed equally to this research.*
[*] *Corresponding author: Yingjie Hu, yhu42@buffalo.edu*

**Abstract**: Social media messages posted by people during natural disasters often contain important location descriptions, such as the locations of victims. Recent research has shown that many of these location descriptions go beyond simple place names, such as city names and street names, and are difficult to extract using typical named entity recognition (NER) tools. While advanced machine learning models could be trained, they require large labeled training datasets that can be time-consuming and labor-intensive to create. In this work, we propose a method that fuses geo-knowledge of location descriptions and a Generative Pre-trained Transformer (GPT) model, such as ChatGPT and GPT-4. The result is a geo-knowledge-guided GPT model that can accurately extract location descriptions from disaster-related social media messages. Also, only 22 training examples encoding geo-knowledge are used in our method. We conduct experiments to compare this method with nine alternative approaches on a dataset of tweets from Hurricane Harvey. Our method demonstrates an over 40% improvement over typically used NER approaches. The experiment results also show that geo-knowledge is indispensable for guiding the behavior of GPT models. The extracted location descriptions can help disaster responders reach victims more quickly and may even save lives.

## 1. Introduction
Natural disasters, such as hurricanes, floods, and tornados, pose significant threats to people and society. Between 2017 and 2021 alone, 89 recorded natural disasters in the United States caused over 4,500 deaths and more than $780 billion in damages and losses (NOAA 2022). With climate change, natural disasters are likely to become even more frequent and more costly in the future

---

unfortunately (Knutson et al. 2010; Elsner, Elsner, and Jagger 2015). Effective disaster response and management are critical for reducing the loss of life and property.

People have made increasing use of social media platforms, such as Twitter and Facebook, during natural disasters to share urgent information and request help (Devaraj, Murthy, and Dontula 2020; J. Wang, Hu, and Joseph 2020; Suwaileh et al. 2022; Zhou et al. 2022). One prominent example is Hurricane Harvey in 2017. A news article published by the U.S. National Public Radio, titled "Facebook, Twitter Replace 911 Calls For Stranded In Houston", reported how affected people used social media to request help and how volunteer responders used those requests to locate and reach the people in need (Silverman 2017). Similar stories were also reported by other news media, such as The Wall Street Journal (Seetharaman and Wells 2017) and Time Magazine (Rhodan 2017). While barriers exist in effectively using social media data for disaster response, a recent survey by Hiltz et al. (2020) showed that emergency managers considered a software system that can automatically process social media data to be "very useful" for disaster management.

Social media messages sent out during natural disasters often contain location descriptions that provide critical geographic information such as the locations of victims and accidents. Figure 1 shows two example tweets posted during Hurricane Harvey (with content slightly modified to protect user privacy) that represent potentially life-or-death scenarios. Accurately extracting these location descriptions and geo-locating them on maps can help disaster responders reach victims more quickly and potentially save lives. While it is possible to recruit many individuals to manually screen these social media messages by hand, a computational method that can automatically and accurately extract these location descriptions can help save time, personpower, and other precious resources during a disaster.

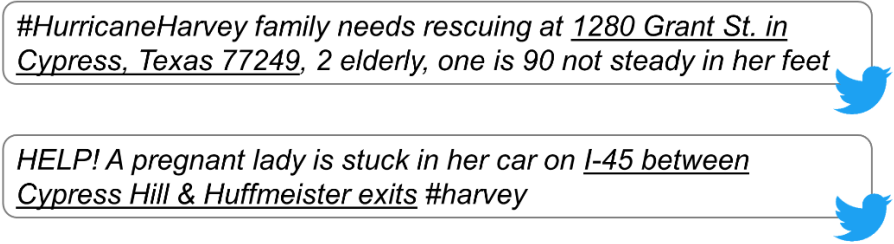

Figure 1. Two rescue-request tweets posted during Hurricane Harvey.

Previous studies have looked into the problem of extracting location descriptions from the content of social media messages (Gelernter and Balaji 2013; Wallgrün et al. 2018; Karimzadeh et al. 2019; J. Wang, Hu, and Joseph 2020; X. Hu et al. 2022; Suwaileh et al. 2022). Two technical steps are typically involved: *recognition* and *geo-locating*. The first step recognizes location descriptions from the textual content of social media messages, while the second step aims to find appropriate geographic coordinates and spatial representations for the recognized location descriptions. We focus on the first step in this work, since recognizing location descriptions is a prerequisite for the second step of geo-locating.

There are two major limitations in previous studies. First, previous studies generally used a default named entity recognition (NER) approach which cannot recognize location descriptions consisting of multiple entities. NER aims to recognize different types of named entities from text, such as *Persons*, *Organizations*, and *Locations*, and it is reasonable to consider the problem of location description recognition as a subtask of NER by simply focusing on *Locations* only. However, recent research has shown that many location descriptions are not in the form of simple place names (e.g., city names or street names) but consist of multiple entities (Y. Hu and Wang 2021; Fernández-Martínez 2022; Chen et al. 2022). Examples of these more complex location descriptions include *door number addresses*, *road intersections*, *highway exits*, and *road segments*. Given that off-the-shelf NER tools are designed to recognize individual entities, they do not have the ability to recognize these multi-entity location descriptions. Figure 2 illustrates this limitation using the two example tweets in Figure 1. Typical NER approaches separately recognize individual entities, such as "Grant St.", "Cypress", and "Texas", rather than the complete location description "1280 Grant St. in Cypress, Texas 77249". This limitation is critical, because our ultimate goal is to properly geo-locate these location descriptions and help first responders reach the people in need; recognizing one complete location description as separate entities can lead to potentially large errors in the geo-locating step, such as locating this location description to the middle of "Grant St." or even to the center of "Texas". From a disaster response perspective, these errors can make first responders arrive at the wrong locations and waste rescue time.

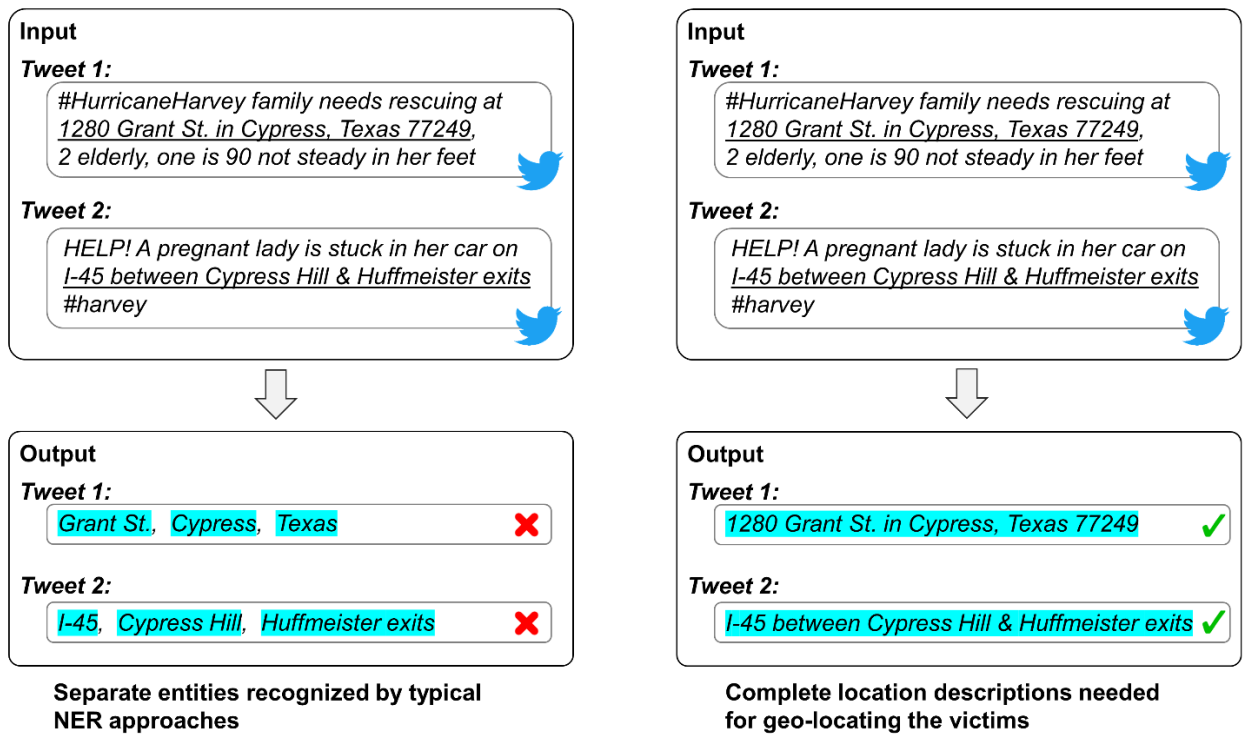

Figure 2. A limitation of typical NER approaches that recognize location descriptions as separate entities rather than complete location descriptions.

A second limitation is that most previous studies have not sought to further classify the recognized location descriptions into more detailed categories, such as *door number addresses*, *road intersections*, and *road segments*. In previous studies, the recognized location descriptions are generally all considered as *locations* and are represented as points on a map (Gelernter and Balaji 2013; Karimzadeh et al. 2019; J. Wang, Hu, and Joseph 2020). Knowing the category of a location description can substantially improve the accuracy of geo-locating, in two ways. First, knowing the categories of location descriptions allows us to choose more suitable geometric representations for them, such as a point for a *road intersection* and a line for a *road segment*. Second, knowing the categories allows us to utilize more suitable geo-locating techniques. For example, we should ideally use the technique of *linear geocoding* if the location description is in the form of *door number addresses* (Goldberg, Wilson, and Knoblock 2007), while we may want to use another technique such as *gazetteer matching* if the location description is in the form of *organization names* (e.g., names of schools and churches). Without knowing the categories of location descriptions, we have to adopt a one-size-fit-all strategy for geo-locating, which inevitably sacrifices the geo-locating accuracy. We note that recognizing complete location descriptions and their categories does not mean we can directly geo-locate them using an existing tool; rather, they facilitate the development of suitable techniques for the second step of geolocating to process the recognized location descriptions in each category. In addition, uncertainty still exists in the identified locations even when we have used suitable geo-locating techniques and geometric representations. It is impossible to measure locations perfectly on the surface of the Earth, and the associated uncertainty issues have long been recognized by researchers (J. Zhang and Goodchild 2002; Goodchild and Haining 2004). For disaster response purposes, a location identified from a description could be considered good enough, if responders, upon arriving at this location, can find the described victims or accidents within a reasonable amount of effort.

The two limitations discussed, i.e., (1) not recognizing complete location descriptions, and (2) not identifying location categories, could theoretically both be overcome by training a new machine learning model on a newly labeled training dataset. This new dataset would require annotating complete location descriptions (rather than separate entities) and also annotating the categories of the location descriptions. This training dataset would also need to be sufficiently large so that the trained machine learning model would achieve satisfactory performance. Creating such a dataset, however, can be time-consuming and labor-intensive.

In recent years, large-scale language models (LLMs) have demonstrated impressive performance in the field of artificial intelligence (AI). LLMs, such as Bidirectional Encoder Representations from Transformers (BERT) (Devlin et al. 2019) and Generative Pre-trained Transformer (GPT) models (Brown et al. 2020; Ouyang et al. 2022), are pre-trained on large-scale textual data (e.g., all text on the Internet) in a task-agnostic manner, and can be adapted to domain-specific tasks via fine tuning, few-shot learning, or sometimes even zero-shot learning. Given their foundational roles in completing various domain-specific tasks, LLMs and other large-scale pre-trained models are also called *foundation models* (Bommasani et al. 2021; Mai et al. 2022).

GPT models, such as GPT-3, ChatGPT, and GPT-4, are LLMs and foundation models that have received substantial attention recently (van Dis et al. 2023). Taking just a few examples as the instruction (called a *prompt*), a GPT model is able to generate text that reads as if it was written by humans. While they are powerful, the current applications of GPT models are largely limited to conversation and text generation, and their social implications are controversial (Dale 2021; Eliot 2022). In this work, we aim to harness the power of GPT models for social good, i.e., to recognize location descriptions from disaster-related social media messages.

A key consideration for harnessing the power of GPT models is the construction of the *prompt*, which serves as the instruction for the model. Since the prompt can be written in a vast number of different ways, how can we create a prompt that makes a GPT model work more effectively for the task of recognizing location descriptions from social media messages? In this work, we propose a geo-knowledge-guided approach for prompt creation, in which the prompt is created based on the geo-knowledge about common forms of location descriptions. The geo-knowledge was obtained and extended from our previous study which systematically examined location descriptions in tweets posted during Hurricane Harvey and identified a set of common forms of location descriptions (Y. Hu and Wang 2021). In this study, we create the prompt based on such geo-knowledge, and feed the created prompt to GPT through a question-answering process. The result is a fusion of the GPT model and geo-knowledge that can overcome the two discussed limitations: it can recognize full location descriptions and also identify the categories of the recognized descriptions. In addition, only a small number of training examples (22 examples in this study) is needed in our method to guide the GPT model.

The remainder of this paper is organized as follows. Section 2 reviews related work on the use of social media for disaster response and location description extraction. Section 3 presents our method that fuses geo-knowledge and GPT models for extracting location descriptions from disaster-related social media messages. Section 4 presents the experiment design for evaluating our method and comparing it with alternative approaches. Section 5 presents the experiment results, and Section 6 discusses the implications of this study on using AI for disaster response. Finally, Section 7 concludes this work.

## 2. Related work

There exists a rich amount of literature on leveraging social media, especially Twitter, for supporting disaster response and situational awareness (De Longueville, Smith, and Luraschi 2009; Starbird and Stamberger 2010; MacEachren et al. 2011; Crooks et al. 2013; Murthy and Longwell 2013; Imran et al. 2015; Imran et al. 2020; Feng, Huang, and Sester 2022). Social media provides near real-time information about the situation on the ground after a disaster (Y. Hu and Wang 2020), which makes it a valuable alternative information source for emergency managers. However, it has been difficult for emergency managers to use the information from social media, due to issues such as large data volume and data veracity (Silverman 2017; Hiltz et al. 2020). Accordingly, much research was devoted to making the information from social media easier to use, including identifying relevant tweets and checking their veracity (Gupta et al. 2013; Joseph,

Landwehr, and Carley 2014; Vosoughi, Roy, and Aral 2018; Imran et al. 2020), classifying the purposes of social media posts (Imran et al. 2014; Yu et al. 2019; Scheele, Yu, and Huang 2021), tracking the transition of different disaster phases (Huang and Xiao 2015; R.-Q. Wang et al. 2020), and monitoring and understanding public sentiments (Ragini, Anand, and Bhaskar 2018; Zou et al. 2018). These are all important research areas, and studies on extracting location descriptions further complement this research landscape.

While geographic locations are considered as highly important by emergency managers (Hiltz et al. 2020), previous studies mostly focused on *geotagged* locations, i.e., locations tagged to tweets (De Albuquerque et al. 2015; Z. Wang, Ye, and Tsou 2016; Martín, Li, and Cutter 2017), rather than locations described in the content of tweets. Recent years have witnessed widespread adoption of social media during disasters to request help and share information (Mihunov et al. 2020; Suwaileh et al. 2022; Zhou et al. 2022). In these social media messages, people describe locations in the content of tweets and may not necessarily geotag their locations. Besides, people may request help for others (such as in the example tweet "*#HurricaneHarvey family needs rescuing at 1280 Grant St. in Cypress, Texas 77249, 2 elderly, one is 90 not steady in her feet*"), and the current location of the Twitter user may not necessarily be the same as the location of the victim. Accordingly, it is critical to extract locations described in the content of social media messages as well.

Previous studies have looked into the problem of extracting locations from the content of social media messages. By considering locations as a special type of named entities, researchers have employed pre-trained NER tools, such as Stanford NER and SpaCy NER, to extract locations from the content of tweets (Gelernter and Balaji 2013; Dutt et al. 2018; Karimzadeh et al. 2019). With the fast advancements of deep learning, researchers have developed deep learning based models for extracting locations. In a previous work, we developed a model called NeuroTPR which improves over a Bidirectional Long Short-Term Memory (BiLSTM) model architecture to extract locations from social media messages (J. Wang, Hu, and Joseph 2020). Given the outstanding performance of transformers more recently, especially BERT, researchers have developed newer methods by leveraging pre-trained or fine-tuned transformers and their variants (X. Hu et al. 2022; Suwaileh et al. 2022; Berragan et al. 2022). While making methodological advancements, previous studies, including our own work, largely considered the problem from a default NER perspective which has two limitations: (1) recognizing individual location entities rather than complete location descriptions; and (2) not classifying location descriptions into more detailed categories. While the NER perspective is good for methodological development, these two limitations constrain our ability to accurately geo-locate victims and accidents during a disaster. It is worth noting that NER tools and models can be retrained using data labeled with complete location descriptions and categories. However, creating such labeled training data requires time, labor, and other resources. There exists another thread of related research on detecting geospatial or, more generally, spatial descriptions from natural language text (Liu, Vasardani, and Baldwin 2014; Stock et al. 2022; Stock, Jones, and Tenbrink 2022). While often studying spatial descriptions under more general contexts (e.g., daily life), research in this thread, such as disambiguating geospatial prepositions (Radke et al. 2022), can be very useful for the step of geo-locating the recognized location descriptions.

The problem of extracting location descriptions from disaster-related social media messages is also related but different from the problem of *geoparsing* in the field of geographic information retrieval (GIR) (Jones and Purves 2008; Freire et al. 2011; Melo and Martins 2017; Purves et al. 2018; Gritta et al. 2018). The goal of *geoparsing* is to recognize and resolve toponyms from texts, such as news articles, Web pages, and social media messages. Geoparsing is typically completed in two steps: toponym recognition and toponym resolution. Existing research in geoparsing often focused on the second step, toponym resolution, to address place name ambiguity issues, while utilizing existing NER tools (e.g., Stanford NER) for the first step toponym recognition (Adelfio and Samet 2013; Y. Hu, Janowicz, and Prasad 2014; DeLozier, Baldridge, and London 2015; Gritta, Pilehvar, and Collier 2018; Karimzadeh et al. 2019; Cardoso, Martins, and Estima 2022). The focus on the second step is reasonable since general geoparsing research often studies ambiguous toponyms that have a world wide coverage (e.g., "Paris" can refer to not only "Paris, France" but also "Paris, Texas"), while the forms of toponyms are often country names and city names (e.g, "Paris") that can be recognized by an NER tool with a relatively high accuracy (J. Wang and Hu 2019b). In comparison, the problem of extracting location descriptions from disaster-related messages presents a different set of challenges, although it can also be completed in a similar two-step process. For the first recognition step, the challenges are in recognizing many of the complex location descriptions consisting of multiple entities (e.g., door number addresses) that cannot be directly recognized by typical NER tools designed for recognizing single entities. Also, it is necessary to further classify location descriptions into more detailed categories, so that suitable geo-locating techniques and geometric representations can be used. For the second step of geo-locating, place name ambiguity becomes less of a concern because we are focusing on a local region affected by the disaster rather than the entire world. However, new challenges for the second step exist in designing and choosing the most suitable geo-locating techniques (e.g., linear geocoding, gazetteer matching, road intersection identification, and highway exit identification) and geometric representations based on the location descriptions and their categories identified in the first step. While different challenges are involved in the problem of location description extraction, the term "toponym" can be defined broadly and the two steps could still be considered as toponym recognition and toponym resolution in a broad sense. In this research, we aim to address the challenges in the first step by exploring a new use of GPT models through the guidance of geo-knowledge.

## 3. Method

Our proposed method fuses geo-knowledge and a GPT model for recognizing complete location descriptions and their categories. This method consists of three components: (1) geo-knowledge about location descriptions, (2) a GPT model, and (3) a process to fuse the two. For (1), we use and extend the geo-knowledge about the common forms of location descriptions obtained from our previous study (Y. Hu and Wang 2021) and encode such knowledge into the prompt. For (2), we employ a pre-trained GPT model. Multiple GPT models, such as GPT-3, ChatGPT, and GPT-4, are tested and studied in this work. For (3), we fuse geo-knowledge and GPT-3 through a question-answering process. Figure 3 provides an overview of our method, and we present methodological details of the three components in the following subsections.

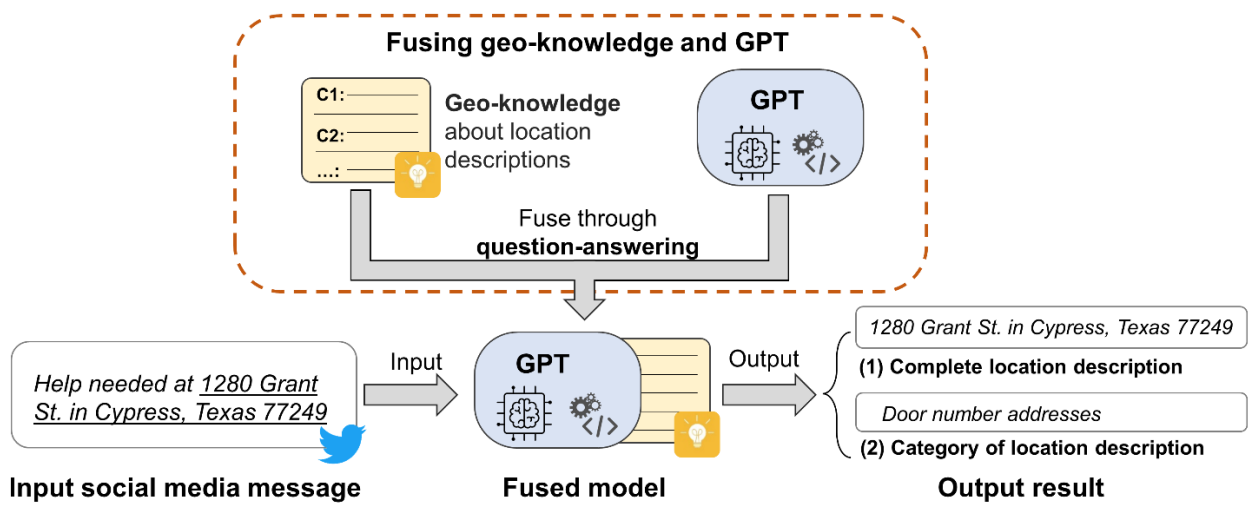

Figure 3. An overview of fusing geo-knowledge and GPT for recognizing location descriptions and their categories from social media messages.

*3.1 Geo-knowledge about location descriptions*

We leverage and extend the geo-knowledge about common forms of location descriptions obtained from our previous study (Y. Hu and Wang 2021). In that work, we assembled a Twitter dataset by randomly selecting 1,000 tweets from a set of 15,834 tweets that are likely to contain location descriptions; the set of 15,834 tweets were selected from a dataset of over 7 million tweets posted during Hurricane Harvey using a regular expression containing location-related terms (e.g., "street", "avenue", "park", "square", "bridge", "rd", and "ave"). We then manually examined each individual tweet and annotated their location descriptions. We also classified these location descriptions into ten categories. Table 1 shows these ten categories (categories *C1* to *C10*) and two example tweets are provided for each category. In this current study, we further extend these categories of location descriptions by adding one more category (category *C11* in Table 1), and will explain why we think such a new category is necessary. We note that a tweet example in Table 1 may contain multiple location descriptions, and we only underscore the location descriptions in a corresponding category in Table 1 for the purpose of clear presentation. All location descriptions in the tweet are annotated in the data.

Table 1: Eleven categories of location descriptions commonly used in Hurricane Harvey extended from (Y. Hu and Wang 2021).

| Category | Location description examples |
|---|---|
| **C1**: Door number addresses | - *"#HurricaneHarvey family needs rescuing at <u>1280 Grant St. in Cypress, Texas 77249</u>, 2 elderly, one is 90 not steady in her feet"* <br> - *"Papa stranded in home. Water rising above waist. HELP <u>812 Wood Ln, 77828</u> #houstonflood"* |
| **C2**: Street names | - *"#Harvey LIVE from San Antonio, TX. Fatal car accident at <u>Ingram Rd.</u>, Strong winds."* <br> - *"<u>Allen Parkway</u>, <u>Memorial</u>, <u>Waugh overpass</u>, Spotts park and Buffalo Bayou park completely under water"* |
| **C3**: Highways | - *"9:00AM update video from Hogan St over White Oak Bayou, <u>I-10</u>, <u>I-45</u>: water down about 4' since last night..."* <br> - *"Left Corpus bout to be in San Angelo #HurricaneHarvey Y'all be safe Avoided <u>highway 37</u> Took the back road"* |
| **C4**: Exits of highways | - *"Need trailers/trucks to move dogs from Park Location: Whites Park Pavillion off I-10 <u>exit 61</u> Anahuac TX"* <br> - *"<u>Townsend exit</u>, Sorters road and Hamblen road is flooded coming from 59 southbound #HurricaneHarvery #Harvey2017"* |
| **C5**: Intersections of roads (rivers) | - *"Guys, this is <u>I-45 @ Main Street</u> in Houston. Crazy. #hurricane #harvey..."* <br> - *"Major flooding at <u>Clay Rd & Queenston</u> in west Houston. Lots of rescues going on for ppl trapped..."* |
| **C6**: Natural features | - *"Frontage Rd at the river #hurricaneHarvey #hurricaneharvey @ <u>San Jacinto River</u>"* <br> - *"<u>Buffalo Bayou</u> holding steady at 10,000 cfs at the gage near Terry Hershey Park"* |
| **C7**: Other human-made features | - *"If you need a place to escape #HurricaneHarvey, <u>The Willie De Leon Civic Center</u>: 300 E. Main St in Uvalde is open as a shelter"* <br> - *"Houston's <u>Buffalo Bayou Park</u> - always among the first to flood. #Harvey"* |
| **C8**: Local organizations | - *"Cleaning supply drive is underway. 9-11 am today at <u>Preston Hollow Presbyterian Church</u>"* <br> - *"#Harvey does anyone know about the flooding conditions around <u>Cypress Ridge High School</u>?! #HurricaneHarvey"* |
| **C9**: Administrative units | - *"#HurricaneHarvey INTENSE eye wall of category 4 Hurricane Harvey from <u>Rockport</u>, <u>TX</u>"* <br> - *"Pictures of downed trees and damaged apartment building on Airline Road in <u>Corpus Christi</u>"* |

| **C10**: Multiple areas | - *"Anyone doing high water rescues in the <u>Pasadena/Deer Park</u> area? My daughter has been stranded in a parking lot all night"* <br> - *"FYI to any of you in <u>NW Houston/Lakewood Forest</u>, Projections are showing Cypress Creek overflowing at Grant Rd"* |
|---|---|
| **C11**: Road segments | - *"HELP! A pregnant lady is stuck in her car on <u>I-45 between Cypress Hill & Huffmeister exits</u> #harvey"* <br> - *"Streets Flooded: <u>Almeda Genoa Rd. from Windmill Lakes Blvd. to Rowlett Rd.</u> #HurricaneHarvey #Houston"* |

The new category that we added is *C11: Road segments*, and an example description in this category is "*HELP! A pregnant lady is stuck in her car on <u>I-45 between Cypress Hill & Huffmeister exits</u> #harvey*". Location descriptions in *C11* were initially put under either category *C4: Exits of highways* or *C5: Intersections of roads (rivers)* in our previous study. While the original classifications are also reasonable (since these location descriptions often involve highway exits and road intersections), we believe that this type of location descriptions may be better represented geometrically as a line in the geo-locating step since they refer to a segment of a road. By contrast, the location descriptions in C4 and C5 are better to be represented as points, since they refer to road intersections or highway exits. With this consideration, we added category C11. The geo-knowledge in this study is therefore the 11 categories of location descriptions and their typical forms represented via examples.

While Table 1 provides one approach for categorizing location descriptions, there exist other approaches and schemes in linguistics and GIScience for organizing locations and more generally spatial information in texts. From a linguistic perspective, Gritta et al. (2020) differentiated *literal toponyms* (e.g., "Paris") and *associative toponyms* (e.g., "Spanish sausages"); Mani et al. (2010) examined both *absolute spatial references* (e.g., "Rome") and *relative spatial references* (e.g., "to the left of the room"); and Pustejovsky et al. (2012) considered the motion of individuals that may be involved in spatial references (e..g, "John biked to Agua Azul"). From a GIScience perspective, one way to classify locations and geographic features is based on their geometric representations, such as points, lines, and polygons (Hill 2000; Longley et al. 2005). The categorization of location descriptions in Table 1 shares some similarities with existing approaches, such as its coverage of absolute spatial references and the use of different geometric representations. Meanwhile, it also adds considerations from a disaster response perspective. First, the categorization in Table 1 is oriented toward location descriptions used under a disaster context. Based on our empirical analysis of tweets from Hurricane Harvey, it seems that people tend to use absolute references (e.g., door number addresses and road intersections) rather than relative or vague references (e.g., to the left of that area) in rescue-request tweets in order to be found by first responders. Second, our categorization considers not only geometric representations but also the different geo-locating techniques necessary to locate these descriptions. For example, door number addresses, road intersections, and highway exits may all be represented as points; however, they require largely different techniques for identifying their locations. Third, our categorization also follows some practices used in previous geoparsing and toponym recognition research in which an NER tool typically differentiates organizations (e.g., schools and churches) from other types of locations and differentiates natural features from human-made features (Gelernter and Mushegian 2011; Karimzadeh et al. 2019).

Nevertheless, the categorization of location descriptions in Table 1 has its limitations. There could be less common location descriptions that are not covered in the current set of categories, which might be uncovered as we analyze more data. In addition, the current categorization is only one version of the geo-knowledge capturing our current understanding of disaster-related location descriptions. As further research is conducted in this area, we will improve our understanding of location descriptions and may refine or even revise the current categorization. One advantage of the proposed method is its flexibility in adjusting the geo-knowledge used to guide the model. When less common location descriptions are identified or when the categorization is improved, we can adjust the geo-knowledge accordingly by changing the examples while still using the same methodological framework.

*3.2 GPT models*

GPT models, such as ChatGPT and GPT 4, have attracted a lot of attention recently from the public media (Eliot 2022; van Dis et al. 2023). GPT models are pre-trained on large-scale textual data, and can generate answers when given a *prompt,* an instruction to the model. Despite their impressive performance in generating human-like text, GPT models have triggered various societal concerns, e.g., they make it easier for students to cheat in homework assignments and may replace certain human jobs that typically require intellectual creativity such as fiction writers (Dale 2021; Kasneci et al. 2023; Mhlanga 2023). In this study, we aim to harness the power of GPT for supporting disaster response by recognizing location descriptions from social media messages.

The behavior of a GPT model is influenced by the prompt it receives. We hypothesize that the best prompt should be created based on systematic knowledge about the target problem, and in this case, it is geo-knowledge about the common forms of location descriptions used by people during natural disasters. To test our hypothesis, we experiment with four different versions of GPT models, which are GPT-2, GPT-3, ChatGPT, and GPT-4. These are all transformer based generative models, and detailed model architecture information is provided in the papers and blog articles by researchers from OpenAI, the company that developed these GPT models (Radford et al. 2018; Radford et al. 2019; Brown et al. 2020; Ouyang et al. 2022; OpenAI 2022; OpenAI 2023). For newer models such as GPT-3, ChatGPT, and GPT-4, OpenAI provides Application Programming Interfaces (APIs) for directly accessing the pre-trained models. For older models such as GPT-2 and GPT, OpenAI does not provide an API but pre-trained models are available from open-source libraries such as the Transformers library from the AI company Hugging Face. We note that while ChatGPT and GPT-4 are very recent models as of April 2023, we started this research back in early 2022 based on GPT-2 and GPT-3; however, given the recent public attention to ChatGPT and GPT-4, we further include them in our experiments. By adding the two latest GPT models, we also demonstrate that our proposed method for fusing geo-knowledge and GPT models is generalizable to newer GPT models which will likely come in the following years.

*3.3 Fusing geo-knowledge and GPT*

We fuse geo-knowledge and GPT by encoding the geo-knowledge of location descriptions into a prompt and feed it to a GPT model to guide its behavior. Learning from previous research on geographic question-answering (Mai et al. 2021), we create the prompt in the form of a series of question-answering statements based on the geo-knowledge of location description categories and their examples shown in Table 1. A snippet of the prompt is provided in Table 2, and we also include the full prompt in the supplementary Table S1 which contains 22 tweet examples in 11 categories (i.e., two examples per category). The prompt first describes the task of location description recognition and the expected output. Then, a series of question-answering examples are provided. Each example is organized as "Sentence", "Q", and "A". The "Sentence" provides a tweet example, "Q" provides a question about the task, and "A" provides the ideal answer that we expect, i.e., complete location descriptions and their categories. In each question-answering example, we "teach" GPT to first predict the category of a location description (e.g., "C1") and then output the full text of the recognized location description. The category and location description are separated by ":". When multiple location descriptions exist in one tweet, we ask the model to separate them by ";". At the end of the prompt, we add a new tweet whose location descriptions are unknown, and we ask the GPT model to infer its location descriptions and their categories based on the previous examples. The final output of GPT is then recorded.

Table 2. A snippet of the prompt with question-answering statements created based on the geo-knowledge of the 11 categories of location descriptions and two examples per category. The full prompt is provided in Supplementary Table S1.

> *This is a set of location description recognition problems.*
> *The `Sentence` is a sentence containing location descriptions.*
> *The goal is to infer which parts of the sentence represent location descriptions and the categories of the location descriptions. Split different location descriptions with `;`.*
> *--*
>
> *--*
> *Sentence: Papa stranded in home. Water rising above waist. HELP 812 Wood Ln, 77828 #houstonflood*
> *Q: Which parts of this sentence represent location descriptions?*
> *A: C1: 812 Wood Ln, 77828*
> *--*
>
> *--*
> *Sentence: Anyone doing high water rescues in the Pasadena/Deer Park area? My daughter has been stranded in a parking lot all night*
> *Q: Which parts of this sentence represent location descriptions?*
> *A: C10: Pasadena/Deer Park*

```
--

--
Sentence: Allen Parkway, Memorial, Waugh overpass, Spotts park and Buffalo Bayou park completely
under water
Q: Which parts of this sentence represent location descriptions?
A: C2: Allen Parkway; C2: Memorial; C2: Waugh overpass; C7: Spotts park; C7: Buffalo Bayou park
--

...

--
Sentence: {TEXT}
Q: Which parts of this sentence represent location descriptions?
A:
```

## 4. Evaluation Experiments

In this section, we describe the evaluation experiments designed to assess the performance of the proposed method. We compare different implementations of our method using different GPT models. We also compare our method with nine alternative approaches, including typically used NER approaches, the traditional fine-tuning approach using BERT, and the default GPT models without the guidance of geo-knowledge. In the following, we present details about the experiment setting.

*4.1 Experiment dataset*

The dataset used for experiments is the 1,000 annotated Hurricane Harvey tweets initially created in our previous work (Y. Hu and Wang 2021) and further extended in this current work. To the best of our knowledge, this is the only dataset available that was annotated with both complete location descriptions and location categories. Since 22 tweets from this dataset have already been used to create the prompt (i.e., those tweet examples in Table 1), we exclude them from testing and use only the remaining 978 tweets to assess the performance of the models. An example of the annotated tweets is shown in Figure 4. The tweets were annotated using the Inside–Outside–Beginning (IOB) tagging scheme (Tjong Kim Sang and De Meulder 2003). The location description category (e.g., "C1" in the figure) was appended to the end of the corresponding "B" and "I" tags. The "O" tag is not explicitly labeled, and all tokens without the "B" or "I" tag are considered as having the "O" tag. The complete dataset is provided in the repository at the end of the article.

> Pls rescue 12 day baby family at **7 Woodview St** , **Houston 77124** , flooding will reach roof soon
>
> B-Location-C1 I-Location-C1 I-Location-C1 I-Location-C1 I-Location-C1

Figure 4. An example of the annotated tweets with location descriptions and their categories labeled using the IOB scheme.

*4.2 Experiment models*

We study our method by experimenting with the following GPT models:

- *Fusing geo-knowledge and GPT-4 (Geo-GPT-4)*: In this implementation, we fuse geo-knowledge and the GPT-4 model using the prompt created based on 22 tweet examples (see Table 2). These tweet examples inform GPT-4 of the forms of location descriptions and the 11 categories of location descriptions with two examples per category.

- *Fusing geo-knowledge and ChatGPT (Geo-ChatGPT)*: This implementation uses the same prompt based on 22 tweet examples as used for *Geo-GPT-4*. Instead of GPT-4, we use ChatGPT for this implementation.

- *Fusing geo-knowledge and GPT-3 (Geo-GPT-3)*: This implementation uses the same prompt but uses the GPT-3 model.

- *Fusing geo-knowledge and GPT-2 (Geo-GPT-2)*: This implementation uses an older GPT model, GPT-2. Because the prompt of this older GPT model cannot be longer than 1024 tokens, we use 11 examples to create the prompt for GPT-2, with one tweet example for each of the 11 categories.

The fusion of geo-knowledge and GPT-3, ChatGPT, and GPT-4 is done through our created prompts and the pre-trained models via the APIs provided by OpenAI. The fusion of geo-knowledge and GPT-2 is done through our created prompt and the pre-trained model from the open-source library from Hugging Face. All source code, such as loading pre-trained models and adapting them through prompts encoding geo-knowledge, is shared in the repository in the Data Availability Statement at the end of the article. In addition to the above four models, we also include nine alternative approaches to serve as baselines:

- *Default GPT-4, ChatGPT, GPT-3, and GPT-2 models:* These default GPT models are included as four baselines to examine the role of geo-knowledge in guiding the behavior of the models. The default GPT models are powerful and were pre-trained with vast amounts of data that could include the same or similar tweets as used in this study. Thus, including these default models can help us understand whether the strong performance of our approach (if any) comes directly from the default GPT models or from the geo-knowledge-guided process. In the prompt to these default models, we ask the same question: "*Which parts of this sentence* *represent location descriptions?*" but do not provide any further geo-knowledge. It is worth noting that the default GPT models do not have the ability to identify location categories, as these categories are part of the geo-knowledge not given to the default models. However, these default models can detect location descriptions based on the large amount of text they have seen during the pre-training process.

- *Fine-tuned BERT (Fine-tuned-BERT)*: Fine-tuning a BERT model is an approach that has been used in recent studies for analyzing disaster-related tweets and has demonstrated good performance (Suwaileh et al. 2022; Zhou et al. 2022). In this baseline, we fine-tune a BERT model using the same 22 tweet examples used to create the prompt for the GPT models. The pre-trained BERT model from Hugging Face is used for implementing this approach.

- *Stanford NER (narrow):* The Stanford NER tool has been commonly used in the literature for recognizing place names from text (Gelernter and Balaji 2013; Liu, Vasardani, and Baldwin 2014; Karimzadeh et al. 2019). Here, we use the off-the-shelf Stanford NER tool (Manning et al. 2014) which can recognize *Person*, *Organization*, and *Location* entities from text. In this *narrow* version of Stanford NER, we keep only *Locations* in the output.

- *Stanford NER (broad):* We use the same off-the-shelf Stanford NER tool but keep both *Locations* and *Organizations* in the output for this *broad* version. The *Locations* output by Stanford NER do not include schools and churches which are considered as *Organizations* by the tool; yet, schools and churches are often used as shelters and their locations are often described during disasters (J. Wang, Hu, and Joseph 2020). Including *Organizations* in the output therefore can help capture those schools and churches. However, this *broad* approach may also include false positives, i.e., organizations that are not used as locations.

- *SpaCy NER (narrow):* SpaCy NER is another tool often used in the literature for recognizing place names from text (Gritta, Pilehvar, and Collier 2018; Y. Hu, Mao, and McKenzie 2019; Fernandes et al. 2021). For this *narrow* version, we use the off-the-shelf SpaCy NER tool and keep only *GPE* (Geopolitical Entities) in the output, which contains cities, counties, states, and countries.

- *SpaCy NER (broad):* We use the same off-the-shelf SpaCy NER tool here but keep both *GPE* (geopolitical entities) and *ORG* (organizations) in the output for this *broad* version. We keep organizations in the output based on the same rationale for the broad version of Stanford NER.

*4.3 Evaluation metrics*

We adopt three evaluation metrics, i.e., *precision*, *recall*, and *F-score*, to assess the performance of the experiment models. These three metrics have been widely used for measuring the performance of computational models in recognizing locations from texts (Gritta et al. 2018; Purves et al. 2018; J. Wang and Hu 2019a). They are calculated using equations (1-3):

$$Precision = \frac{|Correctly\ recognized|}{|All\ recognized|} \quad (1)$$

$$Recall = \frac{|Correctly\ recognized|}{|All\ correct|} \quad (2)$$

$$F-score = 2 \times \frac{Precision \times Recall}{Precision + Recall} \quad (3)$$

*Precision* measures the percentage of correctly recognized location descriptions among all location descriptions recognized by a model. *Recall* measures the percentage of correctly recognized location descriptions among all annotated location descriptions. *F-score* is the harmonic mean of *precision* and *recall*, and *F-score* will be high if both *precision* and *recall* are high and *F-score* will be low if one of the two is low. It is worth noting that the *correctly recognized* location description in Equations (1-2) is measured based on a full-span matching between the model-recognized description and the human-annotated location description. If a recognized location description only partially matches the annotated description, it is considered incorrect. We believe that using full-span matching, rather than allowing partial matching, is important for this research oriented toward disaster response, because a model that recognizes only a part of a location description can lead to limited geo-locating accuracy and even large errors (e.g., erroneously geo-locating a road intersection "Road A & Road B" to the center of "Road A", if only "Road A" is recognized). All experiment models are assessed on the same dataset of 978 tweets (i.e., the 1,000 tweets minus the 22 tweets used for creating the prompt) using full-span matching.

## 5. Results

Two sets of experiments have been conducted to evaluate the performance of the experiment models. In the first set of experiments, we focus on evaluating the recognized location descriptions only, since some of the experiment models (e.g., the default GPT models and the NER models) do not have the ability to recognize the categories of location descriptions. In the second set of experiments, we focus on those models that can recognize both location descriptions and their categories, and evaluate their ability to extract these two types of important information. In the following, we report the results from the two sets of experiments.

*5.1 Ability to recognize complete location descriptions regardless of categories*

The first set of experiments focus on only the text of the recognized location descriptions and does not consider the categories. The results are summarized in Table 3.

Table 3. Precision, recall, and F-score of the experiment models in recognizing complete location descriptions. Metrics are measured based on full-span matching.

| Models | Precision | Recall | F-score |
|---|---|---|---|
| ***Models that cannot identify location categories*** | | | |
| *Stanford NER* (narrow) | 0.621 | 0.402 | 0.488 |
| *Stanford NER* (broad) | 0.564 | 0.440 | 0.495 |
| *SpaCy NER* (narrow) | 0.643 | 0.224 | 0.332 |
| *SpaCy NER* (broad) | 0.352 | 0.340 | 0.346 |
| *GPT-2* | 0.012 | 0.004 | 0.005 |
| *GPT-3* | 0.255 | 0.256 | 0.255 |
| *ChatGPT* | 0.416 | 0.370 | 0.392 |
| *GPT-4* | 0.404 | 0.385 | 0.394 |
| *Models that can identify location categories* | | | |
| *Geo-GPT-2* | 0.380 | 0.404 | 0.391 |
| *Geo-GPT-3* | **0.693** | 0.694 | 0.693 |
| *Geo-ChatGPT* | 0.633 | 0.673 | 0.653 |
| *Geo-GPT-4* | 0.687 | **0.704** | **0.695** |
| *Fine-tuned-BERT* | 0.150 | 0.242 | 0.185 |

We first look at the performance of the NER models in Table 3. The four NER models achieve precisions between 0.352 and 0.643, recalls between 0.224 and 0.440, and F-scores between 0.332 and 0.495. While these results are not disappointing, a closer examination shows that the recognized location descriptions are all in the form of simple place names, such as city names, state names, and river names. In fact, they completely miss those more complex location descriptions, such as door number addresses, road segments, and road intersections, which consist of multiple entities. This result is unsurprising, as we know that these NER models are designed to recognize individual named entities and cannot recognize location descriptions consisting of multiple entities. Meanwhile, these multi-entity location descriptions provide highly detailed location information that can help rescue teams to reach victims. Models that fail to recognize these detailed location descriptions can provide only limited support for disaster response efforts.

Next, we look at the performance of the default GPT models and the Geo-GPT models. As shown in Table 3, there is an increase in performance from GPT-2 to GPT-4, demonstrating that these GPT models are indeed becoming more intelligent over the years. While GPT-2 mostly fails in the experiment, GPT-3, ChatGPT, and GPT-4 are able to recognize about 25%-40% of the location descriptions, including some more complex multi-entity location descriptions, such as door number addresses and road intersections. However, the performance of these default GPT models is still quite limited, with the highest F-score 0.394 achieved by GPT-4. These default GPT models also do not have the ability to identify the categories of location descriptions. In comparison, the Geo-GPT models substantially improve over their corresponding default GPT models in all metrics. This improvement can be seen across all four GPT models from GPT-2 to GPT-4, demonstrating that the effectiveness of geo-knowledge in improving performance is not limited to one particular model but applicable to all GPT models. Geo-knowledge plays an important role by informing the GPT models about the common forms of location descriptions and the categories of these descriptions. Without geo-knowledge, the default GPT models can rely on only the information of location descriptions that they obtained during the pre-training process, which likely leads to lower performance. While geo-knowledge is important, the advancement in GPT models is also necessary for achieving the obtained good results. As can be seen in Table 3, the model performance also increases from Geo-GPT-2 to Geo-GPT-4 in general. Interestingly, Geo-ChatGPT shows a slightly lower performance than that of Geo-GPT-3, even though ChatGPT is a newer model than GPT-3. Finally, the fine-tuned BERT model achieves a precision 0.150, a recall 0.242, and an F-score 0.185. These low performance scores are likely due to the small number of training examples not sufficient for adapting a complex model like BERT.

The highest F-score of 0.695 is achieved by Geo-GPT-4, and a similar score of 0.693 is achieved by Geo-GPT-3. These F-scores are close to the threshold of 0.70 considered as acceptable in location recognition tasks for disaster response (Suwaileh et al. 2022). While these scores are probably not outstanding, they show an over 40% improvement compared with off-the-shelf NER models, such as the Stanford NER used in the experiments. NER models can be re-trained to achieve a similar performance using a sufficiently large dataset labeled with complete location descriptions and categories; however, creating such a dataset requires time, labor, and other precious resources. The geo-knowledge-guided GPT models therefore present an efficient approach for extracting location descriptions when we have only a small number of training data examples encoding geo-knowledge. To further examine the sensitivity of Geo-GPT-4 to different examples included in the prompt, we test two additional prompts, one with a different set of 22 tweets and the other with a set of 22 tweets synthesized based on the geo-knowledge. The performance of Geo-GPT-4 based on these two prompts slightly increases (an increase of 0.023 and 0.008 in terms of F-score), but does not substantially change. These two prompts and the test results are included in Supplementary Tables S2-S4.

*5.2 Ability to recognize both complete location descriptions and location categories*
In the second set of experiments, we evaluate the ability of the models to recognize both complete location descriptions and location categories. We focus on the models that have the ability to recognize location categories in addition to the description text, which are Geo-GPT-2, Geo-GPT-3, Geo-ChatGPT, Geo-GPT-4, and fine-tuned BERT. A recognized location description is considered correct only when it has both the complete location description and the correct location category. We compute the precision, recall, and F-score of the five models for each of the 11 categories and their overall scores across all categories, and the results are summarized in Figure 5.

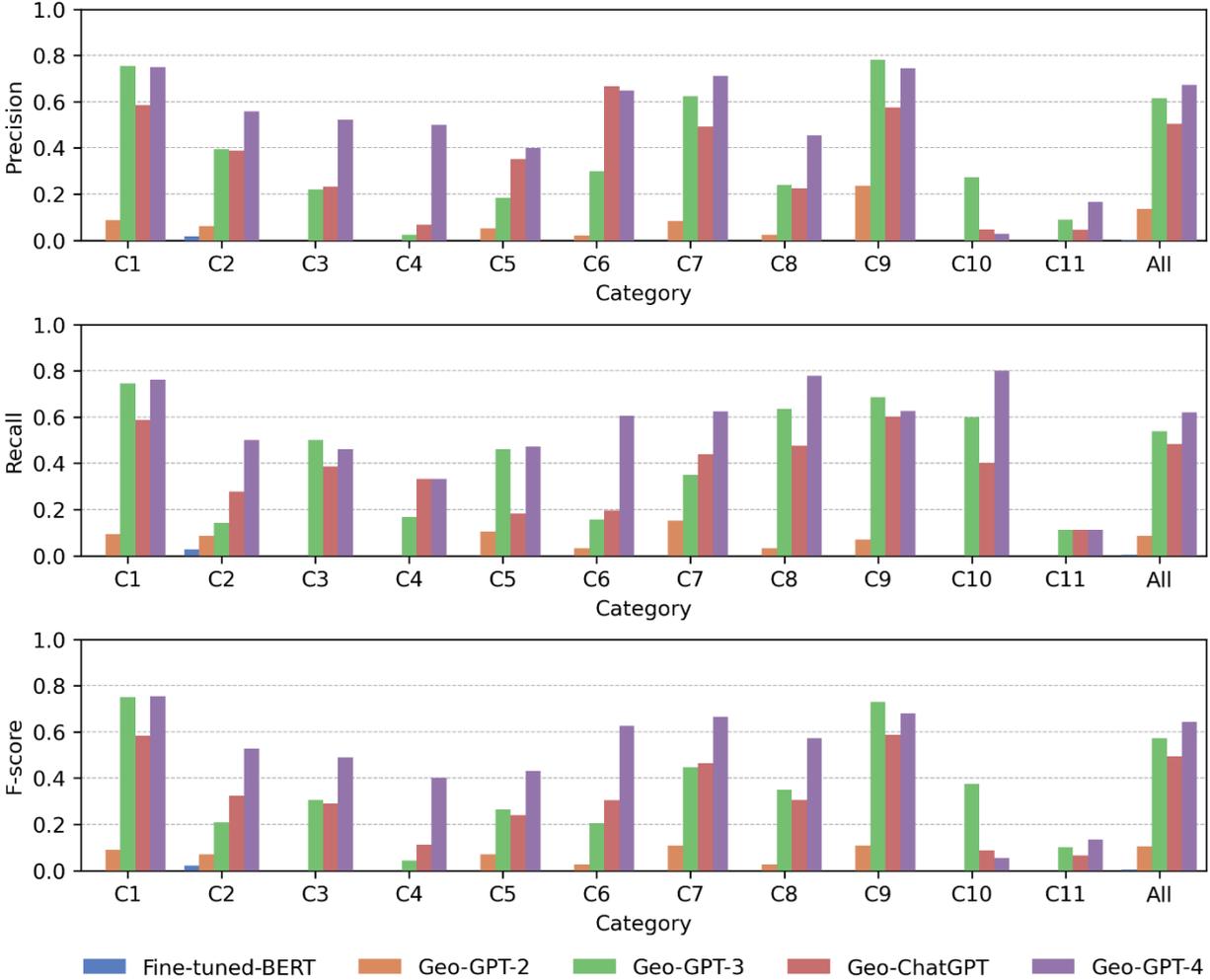

Figure 5. Precision, recall, and F-score of the tested models in recognizing both complete location descriptions and their categories.

As shown in the figure, Geo-GPT-3, Geo-ChatGPT, and Geo-GPT-4 achieve higher performance scores than the other two models in each of the eleven location description categories and also across all categories. The fine-tuned BERT model fails to identify the correct category of a location description most of the time, even in cases when it recognizes the full location description. This result again is likely due to the fact that the small number of training examples are not sufficient to adapt the BERT model, especially for this relatively complex problem with eleven categories to be identified. While Geo-GPT-2 improves over the fine-tuned BERT model, its performance is still quite limited and its precisions, recalls, and F-scores in most location categories are lower than 0.2. Substantial performance improvements are observed in Geo-GPT-3, Geo-ChatGPT, and Geo-GPT-4. Consistent with the first set of experiments, Geo-GPT-3 and Geo-GPT-4 perform better than Geo-ChatGPT in most categories and overall, although Geo-ChatGPT does perform better than Geo-GPT-3 in some categories, such as *C2: Street names* and *C6: Natural features*. It is probably less surprising that Geo-GPT-4 shows a better performance than Geo-ChatGPT, given that GPT-4 is a newer model than ChatGPT and OpenAI has shown that GPT-4 outperformed ChatGPT in multiple tests, such as bar exams (OpenAI 2023). It does surprise us that Geo-GPT-3 performs better than Geo-ChatGPT. According to OpenAI, ChatGPT has been further fine-tuned using a new dialogue dataset to optimize its performance toward chat-related tasks (OpenAI 2022). Such chat-related optimization might have affected the performance of ChatGPT on this current location description recognition task. Both Geo-GPT-3 and Geo-GPT-4 achieve good performance in category *C1: Door number addresses,* with precisions, recalls, and F-scores all about 0.75. Also, both models extract whole door number addresses including the prepositions (e.g., "in") sometimes used by people but not typically part of formal addresses. These extracted whole addresses allow us to analyze and geo-locate them in the next step by building on previous research, such as research on geospatial preposition analysis (Radke et al. 2022). Given that door number addresses provide precise location information, correctly recognizing location descriptions in this category can be highly helpful in identifying the locations of victims. Geo-GPT-4 demonstrates better performance than Geo-GPT-3 over the majority of the categories and also across all categories, with an overall F-score of 0.644 achieved by Geo-GPT-4 and an overall F-score of 0.573 achieved by Geo-GPT-3. This result helps us further understand the performance difference between Geo-GPT-3 and Geo-GPT-4: while the two show similar performance in the first set of experiments focusing on location descriptions only, Geo-GPT-4 demonstrates a substantially better performance in correctly identifying both location descriptions and their categories.

To understand the errors made by Geo-GPT-4, we create a confusion matrix in Figure 6. In this figure, each row represents location descriptions in a category in the test data (i.e., ground truth), and each column represents location descriptions recognized by Geo-GPT-4 in that category (i.e., model predictions). The integer number in a cell represents the number of location descriptions correctly recognized by the model. For example, the number "200" at row C1 and column C1 indicates that 200 of the door number addresses in the ground truth were correctly recognized by Geo-GPT-4; meanwhile, the number "2" at row C1 and column C9 indicates that one door number address was correctly recognized by Geo-GPT-4 but was mistakenly classified as category C9. The percentage value in each cell is calculated based on the row sum. For example, the value "76.05%" at row C1 and column C1 indicates that the 200 corrected recognized location descriptions represent 76.05% of all labeled location descriptions in category C1. There are also situations in which labeled location descriptions are not correctly recognized or completely missed by the model, which are represented in the "Missed" column; there also exist recognized location descriptions that are not in the ground truth (i.e., false positives), which are represented in the "Not in ground truth" row.

|  | C1 | C2 | C3 | C4 | C5 | C6 | C7 | C8 | C9 | C10 | C11 | Missed |
|---|---|---|---|---|---|---|---|---|---|---|---|---|
| Door numbers: C1 | **200** **76.05%** | 0 0.00% | 0 0.00% | 0 0.00% | 0 0.00% | 0 0.00% | 0 0.00% | 0 0.00% | 2 0.76% | 0 0.00% | 0 0.00% | 61 23.19% |
| Street names: C2 | 0 0.00% | **112** **50.00%** | 0 0.00% | 0 0.00% | 6 2.68% | 0 0.00% | 0 0.00% | 0 0.00% | 1 0.45% | 0 0.00% | 0 0.00% | 105 46.88% |
| Highways: C3 | 0 0.00% | 0 0.00% | **12** **46.15%** | 0 0.00% | 0 0.00% | 0 0.00% | 0 0.00% | 0 0.00% | 0 0.00% | 0 0.00% | 0 0.00% | 14 53.85% |
| Exits of highways: C4 | 0 0.00% | 0 0.00% | 0 0.00% | **2** **33.33%** | 0 0.00% | 0 0.00% | 0 0.00% | 0 0.00% | 0 0.00% | 0 0.00% | 0 0.00% | 4 66.67% |
| Intersections: C5 | 0 0.00% | 5 4.81% | 1 0.96% | 0 0.00% | **49** **47.12%** | 2 1.92% | 0 0.00% | 0 0.00% | 0 0.00% | 0 0.00% | 0 0.00% | 47 45.19% |
| Natural features: C6 | 0 0.00% | 0 0.00% | 0 0.00% | 0 0.00% | 0 0.00% | **37** **60.66%** | 1 1.64% | 0 0.00% | 3 4.92% | 11 18.03% | 0 0.00% | 9 14.75% |
| Other human-made: C7 | 0 0.00% | 12 5.38% | 0 0.00% | 0 0.00% | 0 0.00% | 3 1.35% | **139** **62.33%** | 9 4.04% | 3 1.35% | 19 8.52% | 0 0.00% | 38 17.04% |
| Local organizations: C8 | 0 0.00% | 0 0.00% | 0 0.00% | 0 0.00% | 0 0.00% | 0 0.00% | 1 1.59% | **49** **77.78%** | 0 0.00% | 0 0.00% | 0 0.00% | 13 20.63% |
| Admin units: C9 | 0 0.00% | 0 0.00% | 0 0.00% | 0 0.00% | 3 0.42% | 0 0.00% | 2 0.28% | 4 0.56% | **444** **62.54%** | 49 6.90% | 0 0.00% | 208 29.30% |
| Multiple areas: C10 | 0 0.00% | 0 0.00% | 0 0.00% | 0 0.00% | 0 0.00% | 0 0.00% | 0 0.00% | 0 0.00% | 0 0.00% | **4** **80.00%** | 0 0.00% | 1 20.00% |
| Road segments: C11 | 0 0.00% | 2 22.22% | 0 0.00% | 0 0.00% | 3 33.33% | 0 0.00% | 0 0.00% | 0 0.00% | 0 0.00% | 0 0.00% | **1** **11.11%** | 3 33.33% |
| Not in ground truth | 67 12.38% | 70 12.94% | 10 1.85% | 2 0.37% | 62 11.46% | 15 2.77% | 52 9.61% | 46 8.50% | 143 26.43% | 58 10.72% | 5 0.92% | 11 2.03% |

Figure 6. Confusion matrix based on the output of Geo-GPT-4 and the annotated location descriptions.

The confusion matrix helps us further understand the performance of Geo-GPT-4 across different categories of location descriptions. It is worth noting that some categories of location descriptions, such as C1, C2, C7 and C9, show up more frequently than some others in the data. However, we cannot focus on only those more frequent location descriptions, for two reasons. First, some categories of location descriptions that show up more frequently are probably less informative from the perspective of disaster response. One example is *C9: Administrative units*: while city and state names, such as "Houston" and "Texas", show up frequently in the data, they are less useful for first responders to locate and reach the victims. Second, location descriptions that show up less frequently are still used in multiple tweets describing life-threatening situations. Recognizing these relatively less frequent location descriptions can be critical for reaching the people in need. Overall, Geo-GPT-4 is effective in correctly recognizing both complete location descriptions and the categories of the descriptions, as can be seen in the diagonal of the confusion matrix. Across different categories, Geo-GPT-4 makes a small number of errors in which a correctly recognized location description is mistakenly classified into another category. Geo-GPT-4 does make more mistakes in category *C11: Road segments* by misclassifying them to *C2: Street names* and *C5: Intersections*. A description of a road segment, such as "*Almeda Genoa Rd. from Windmill Lakes Blvd. to Rowlett Rd.*", does often involve street names and road intersections, which could be a reason for the confusion made by Geo-GPT-4.

The majority of the errors fall into either the column of "Missed" or the row of "Not in ground truth". By examining these errors based on the output of Geo-GPT-4, we find that quite some location descriptions are in fact reasonably recognized by the model; however, they do not completely match the annotated location description given our requirement of strict full-span matching in the evaluation experiments. For example, in the tweet "*#Houston #HoustonFlood This is the intersection of I-45 & N. Main Street*", the recognized location description of Geo-GPT-4 is "*the intersection of I-45 & N. Main Street*", while the annotated description is "*I-45 & N. Main Street*". We note that a clean string of "*I-45 & N. Main Street*" is easier for the next step of geo-locating, but the recognized description "*the intersection of I-45 & N. Main Street*" may not be considered as completely wrong either. However, in this example, the ground truth annotation "*I-45 & N. Main Street*" is counted as "Missed", while the recognized description "*the intersection of I-45 & N. Main Street*" is counted as "Not in ground truth". There exist other similar cases, e.g., the output of the model is "*Sugarland area*" while the annotation is "*Sugarland*". Sometimes, there is a larger difference between the recognized location description and the annotated description. For example, in the tweet "*Spokeswoman for Houston Mayor Sylvester Turner says the convention center at NRG Park is opening, serving 10,000 additional Harvey evacuee*", the annotated location description is "*NRG Park*" while the description recognized by the model is "*convention center at NRG Park*". Again, the model output may not be considered as completely wrong, although it is different from the annotated location description. These and other similar cases have contributed to the relatively larger errors observed in the "Missed" column and the "Not in ground truth" row in the confusion matrix. If we consider those reasonable location descriptions as correct using a relaxed matching method (Li et al. 2020), the performance of Geo-GPT-4 could be higher than the obtained scores. In Supplementary Figure S1, we show another confusion matrix which is based on the same result of Geo-GPT-4 but uses a relaxed matching method that allows the recognized location descriptions that have over 75% overlapping with the annotated descriptions to be considered as correct. This relaxed confusion matrix shows a 29.8% decrease of the errors in the "Missed" column and a 28.3% decrease of the errors in the "Not in ground truth" row, compared with the confusion matrix in Figure 6. Most of these partially matched descriptions, initially considered as errors, are also classified under the right location description categories, although some of them are still misclassified. While this result is encouraging, we need to be careful about using partially matched location descriptions as they could lead to potentially large errors in the geo-locating step.

# 6. Discussion

*6.1 The importance of both geo-knowledge and GPT models for the proposed method*

In this work, we have proposed a method that fuses geo-knowledge and a GPT model for extracting location descriptions and their categories from disaster-related social media messages. Both geo-knowledge and GPT models are important for the proposed method. Without geo-knowledge, the default GPT models do not know the typical forms of location descriptions used by people during disasters and how to categorize these location descriptions. As a result, GPT models have to rely on the information of location descriptions that they obtained during the pre-training stage. As demonstrated in our experiment result, adding geo-knowledge to the GPT models increases their performance by over 76%, and enables the GPT models to identify the categories of location descriptions. Meanwhile, advancements in GPT models are important for our proposed method as well. Without the invention of the GPT models, we cannot effectively integrate the geo-knowledge represented by a small number of examples into a typical machine learning model. In addition, we also observe an increased performance overall when a more advanced GPT model is used. In sum, both geo-knowledge and GPT models are indispensable for the proposed method, and fusing the two enables us to effectively recognize location descriptions and their categories from disaster-related social media messages.

*6.2 Implications for research in other geographic regions and data from other platforms*

While this study has used tweets from Hurricane Harvey in the Houston area in Texas, the proposed method has flexibility and we expect that it could be applied to other geographic regions. For research within the United States, our method could be used directly since the geo-knowledge is still applicable. The location descriptions and their categories, such as door number address, road intersections, and highways, are commonly used in other geographic regions in the U.S., and are likely to be used by people in future disasters within the country. For research in other countries, the proposed method could be used by replacing the geo-knowledge of U.S. location descriptions with the geo-knowledge of the specific local region. In cases when the geo-knowledge of a local region is not available, research may need to be conducted to identify the local location descriptions, organize them into examples, and use the examples to guide GPT models in the same way as done in this research. In addition, since GPT models can handle multilingual texts, the proposed method could be used to analyze text messages in other languages beyond English. Nevertheless, future investigations are necessary to empirically test the applicability of the proposed method in other geographic regions and languages.

Our method may also be used to analyze text messages from other platforms, not limited to Twitter. There exists uncertainty related to the recent ownership change of Twitter and the announced changes in Twitter data API. While the future of this social media company may be unclear, it seems that people do need a digital platform during a disaster to post information and learn current situations (Pourebrahim et al. 2019; Mihunov et al. 2020; C. Zhang, Yang, and Mostafavi 2021). If Twitter engagement were to drop significantly, people may switch to other social media platforms, or a new digital platform for disaster response may be developed such as the Ushahidi platform used in the Haiti Earthquake in 2010 (Meier 2010). Our proposed method uses only the textual content of tweets, and does not rely on other data or features that are specific to Twitter. Therefore, it could also be used for extracting location descriptions and their categories from text messages from other platforms.

*6.3 A potential paradigm shift for using AI models for disaster response?*
AI models have been increasingly adopted in disaster response (Kuglitsch et al. 2022). A typical approach in most recent studies is to fine-tune a pre-trained AI model, such as BERT, using newly labeled data (J. Wang, Hu, and Joseph 2020; Suwaileh et al. 2022; Zhou et al. 2022). While such an approach avoids training an AI model completely from scratch, it still requires hundreds or thousands of annotated examples depending on the complexity of the specific task. In addition to training data, configuring the deep learning computing environment and fine-tuning AI models put additional requirements on disaster response organizations in terms of their technical expertise and computing resources. As shown in this study, fusing geo-knowledge and GPT models has the potential to substantially simplify the adoption of an AI model to a disaster response task.

On the one hand, disaster response organizations do not need to fine-tune their own AI models and configure their own deep learning environments locally. Instead, they can leverage the online AI models, such as the GPT models used in this study, and formalize their knowledge and other information about the task into prompts for adopting the AI model. It seems that such an approach can largely reduce the requirements of training data, technical expertise, computing resources (both hardware and software), and time put on disaster response organizations. In addition, this approach may also facilitate the collaboration between disaster response experts and AI model developers by allowing them to focus on what they are mostly good at: disaster response experts can focus on obtaining and formalizing knowledge by analyzing disaster-related data, while model developers can focus on developing and improving AI models. The knowledge obtained by disaster response experts can then be fused with the AI model hosted on the cloud computing environment through an API using prompts, similar to what we did in this study. Through such a collaboration, we may be able to empower disaster response organizations to utilize AI models more easily and more efficiently, and let them focus on reaching and helping the people in need during a disaster.

On the other hand, there also exist important ethical issues that deserve our careful consideration. In the proposed approach, the social media messages of the disaster victims need to be submitted to an online GPT model, and malicious actors could take advantage of those submitted messages. While one could argue that those messages were shared as public tweets and could still be used by bad actors in their original form, further submitting them to an online model can increase their exposure. At the time of writing, a prompt of "What were the addresses of the people who needed help during Hurricane Harvey?" tested on ChatGPT was not replied with any address, and ChatGPT included a statement in its reply that "sharing personal addresses or identifying information would violate privacy and confidentiality guidelines." Also, according to the Terms of Use from OpenAI, the content submitted via their API is not used to develop or improve their models. While it is good to see these efforts, further policies and technical measures are necessary to safeguard the content submitted to AI models. In addition, similar to other online platforms, who owns the data that are submitted to an online AI model? If law enforcement requests to use such data as evidence, do they have the right to do so? Should we apply stricter data protection policies for the content submitted under urgent situations such as natural disasters than that submitted during normal time? These are some of the important questions that we need to answer in order to use large AI models for disaster response in an ethical manner.

Could there be a paradigm shift in using AI models for disaster response, i.e., from locally trained and deployed AI models to online AI models guided by knowledge? While our work has shown promising results, more studies are necessary to understand the pros, cons, and ethical issues of using large online AI models for disaster response.

*6.4 Limitations*

This research is not without limitations. First, the performance of the Geo-GPT-4 model is still low on some location description categories, such as *C11: Road segments*. Based on our analysis of the experiment results, it seems that the model often confuses C11 with *C2: Street names* and *C5: Intersections*. While C11 indeed shares some similarities with the other two categories, we may need to think of better ways to inform GPT models about the differences among these categories. Second, our current study has focused on a dataset from the Houston area from Hurricane Harvey. While our method has potential to be applied to data from other geographic regions and other types of disasters, more empirical research is necessary to test the wider applicability of this method. Third, the overall performance of the Geo-GPT-4 model is still limited and could be improved through new methodological development in the near future. For example, we have used a question-answering approach in this study to encode geo-knowledge for informing GPT models, and future research could explore other ways to encode geo-knowledge, not limited to a series of questions and answers. Finally, this current study has focused on the step of recognizing location descriptions only, and it is important to explore the second step of geo-locating these descriptions as well. With the recognized location descriptions and their categories, we can further analyze and geo-locate them using suitable techniques and geometric representations. As mentioned earlier, uncertainty still exists even when suitable geo-locating techniques and geometric representations are used. A Bayesian approach could be adopted to quantify such uncertainty based on the location information contained in a message. For example, city name alone may be associated with a higher location uncertainty, while additional information, such as street name and door number, helps reduce such uncertainty.

## 7. Conclusions

Social media platforms are increasingly being used by people to share information and request help during natural disasters. Messages posted on these platforms often contain important descriptions about the locations of victims and accidents. In this study, we proposed a method that fuses geo-knowledge and GPT models for recognizing location descriptions and their categories.

We presented the methodological details and conducted systematic experiments to compare the proposed method with other alternative approaches, including the typically used NER approaches and default GPT models. We found that geo-knowledge-guided GPT models achieve an over 40% improvement in recognizing location descriptions compared with off-the-shelf NER approaches, and this method uses only a small number of training examples encoding geo-knowledge. We also found that both geo-knowledge and GPT models are critical for the proposed method. Adding geo-knowledge to a GPT model results in an over 76% improvement compared with the same default GPT model; meanwhile, there is a general increase in performance when a more advanced GPT model is used. The highest performance was achieved by Geo-GPT-4 which demonstrated an F-score of 0.695 for recognizing location descriptions regardless of categories and an overall F-score of 0.644 for recognizing both location descriptions and their categories. In addition, Geo-GPT-4 achieved an F-score of 0.755 for the category of door number address which provides highly detailed information for locating victims. This method has been tested with different versions of GPT models, including GPT-2, GPT-3, ChatGPT, and GPT-4, and is likely to be applicable to more advanced GPT models in the coming years. By fusing geo-knowledge and GPT models, we may facilitate collaborations between disaster response experts and AI developers, reduce the technical burdens on disaster response organizations, and ultimately help save lives.

**Acknowledgement**
We thank the anonymous reviewers for their constructive comments and suggestions.

**Disclosure statement**
No potential conflict of interest was reported by the authors.

**Funding**
This work is supported by the U.S. National Science Foundation under Grant No. BCS-2117771: "Geospatial Artificial Intelligence Approaches for Understanding Location Descriptions in Natural Disasters and Their Spatial Biases." Any opinions, findings, and conclusions or recommendations expressed in this material are those of the authors and do not necessarily reflect the views of the National Science Foundation.

**Notes on contributors**
**Yingjie Hu** is an Associate Professor in the Department of Geography at the University at Buffalo. His research interests include GIScience, geospatial artificial intelligence (GeoAI), and disaster resilience. His contributions to this paper include conceptualization, data collection and curation, methodology, data analysis, result interpretation and discussion, visualization, writing – original draft and writing – editing and revision.

**Gengchen Mai** is currently a Tenure-Track Assistant Professor in the Department of Geography at the University of Georgia. His research interests are spatially explicit artificial intelligence, geographic knowledge graphs, geographic question answering, geospatial foundation models, and so on. His contributions to this paper include conceptualization, methodology, data analysis, result interpretation and discussion, visualization, writing – original draft, and writing – editing and revision.

**Chris Cundy** is a Ph.D. student at Stanford University. His research interests include variational inference, generative models, large language models, and AI safety. His contributions to this paper include methodology, result interpretation and discussion, and writing – editing and revision.

**Kristy Choi** is a Ph.D. candidate in Computer Science at Stanford University advised by Dr. Stefano Ermon. Her research is centered around machine learning with limited labeled supervision, and is focused on developing techniques for better adaptation and controllability in deep generative models. Her contributions to this paper include methodology, result interpretation and discussion, and writing – editing and revision.

**Ni Lao** is currently a research scientist at Google. He holds a Ph.D. degree in Computer Science from the Language Technologies Institute, School of Computer Science at Carnegie Mellon University, and is an expert in machine learning, knowledge graph and natural language understanding. His contributions to this paper include methodology, result interpretation and discussion, and writing – editing and revision.

**Wei Liu** has recently graduated with a M.S. in GIS from the University at Buffalo. Her interests include GeoAI, urban analytics, and social equality. Her contributions to this paper include data collection and curation and data analysis.

**Gaurish Lakhanpal** recently graduated from Stevenson High School and is currently studying computer science at Purdue University. He is deeply interested in machine learning. His contributions to this paper include data collection and curation, data analysis, and visualization.

**Ryan Zhenqi Zhou (Zhenqi Zhou)** is a Ph.D. candidate in the Department of Geography at the University at Buffalo and a research assistant in the GeoAI Lab. His research interests include GeoAI, disaster resilience, public health, and human mobility. His contributions to this paper include data collection and curation and data analysis.

**Kenneth Joseph** is an Assistant Professor in the Computer Science and Engineering Department at the University at Buffalo. He is a computational social scientist who focuses on the measurement and modeling of complex social systems. His contributions to this paper include result interpretation and discussion and writing – editing and revision.

**Data and code availability statement**
The data and code that support the findings of this study are available on figshare at https://doi.org/10.6084/m9.figshare.22659337.

*Geographical Information Science* 29 (4): 667–689.
De Longueville, Bertrand, Robin S Smith, and Gianluca Luraschi. 2009. "' OMG, from Here, I Can See the Flames!' A Use Case of Mining Location Based Social Networks to Acquire Spatio-Temporal Data on Forest Fires." In , 73–80.
DeLozier, Grant, Jason Baldridge, and Loretta London. 2015. "Gazetteer-Independent Toponym Resolution Using Geographic Word Profiles." In *Twenty-Ninth AAAI Conference on Artificial Intelligence*.
Devaraj, Ashwin, Dhiraj Murthy, and Aman Dontula. 2020. "Machine-Learning Methods for Identifying Social Media-Based Requests for Urgent Help during Hurricanes." *International Journal of Disaster Risk Reduction* 51. Elsevier: 101757.
Devlin, Jacob, Ming-Wei Chang, Kenton Lee, and Kristina Toutanova. 2019. "BERT: Pre-Training of Deep Bidirectional Transformers for Language Understanding." In , 4171–4186. Minneapolis, Minnesota: Association for Computational Linguistics.
Dutt, Ritam, Kaustubh Hiware, Avijit Ghosh, and Rameshwar Bhaskaran. 2018. "Savitr: A System for Real-Time Location Extraction from Microblogs during Emergencies." In *Companion Proceedings of the The Web Conference 2018*, 1643–1649.
Eliot, Lance. 2022. "Enraged Worries That Generative AI ChatGPT Spurs Students To Vastly Cheat When Writing Essays, Spawns Spellbound Attention For AI Ethics And AI Law." *Forbes*. https://www.forbes.com/sites/lanceeliot/2022/12/18/enraged-worries-that-generative-ai-chatgpt-spurs-students-to-vastly-cheat-when-writing-essays-spawns-spellbound-attention-for-ai-ethics-and-ai-law.
Elsner, James B., Svetoslava C. Elsner, and Thomas H. Jagger. 2015. "The Increasing Efficiency of Tornado Days in the United States." *Climate Dynamics* 45. Springer: 651–659.
Feng, Yu, Xiao Huang, and Monika Sester. 2022. "Extraction and Analysis of Natural Disaster-Related VGI from Social Media: Review, Opportunities and Challenges." *International Journal of Geographical Information Science* 36 (7). Taylor & Francis: 1275–1316.
Fernandes, Chelsea, Joshua Fernandes, Sharon Mathew, Shubham Raorane, and Anuradha Srinivasaraghavan. 2021. "Automated Disaster News Collection Classification and Geoparsing." In *Proceedings of the International Conference on Smart Data Intelligence (ICSMDI 2021)*.
Fernández-Martínez, Nicolás José. 2022. "The FGLOCTweet Corpus: An English Tweet-Based Corpus for Fine-Grained Location-Detection Tasks." *Research in Corpus Linguistics* 10 (1): 117–133.
Freire, Nuno, José Borbinha, Pável Calado, and Bruno Martins. 2011. "A Metadata Geoparsing System for Place Name Recognition and Resolution in Metadata Records." In *Proceedings of the 11th Annual International ACM/IEEE Joint Conference on Digital Libraries*, 339–348.
Gelernter, Judith, and Shilpa Balaji. 2013. "An Algorithm for Local Geoparsing of Microtext." *GeoInformatica* 17 (4). Springer: 635–667.
Gelernter, Judith, and Nikolai Mushegian. 2011. "Geo-parsing Messages from Microtext." *Transactions in GIS* 15 (6): 753–773.
Goldberg, Daniel W., John P. Wilson, and Craig A. Knoblock. 2007. "From Text to Geographic Coordinates: The Current State of Geocoding." *URISA Journal* 19 (1). Citeseer: 33–46.
Goodchild, Michael F., and Robert P. Haining. 2004. "GIS and Spatial Data Analysis: Converging Perspectives." *Papers in Regional Science* 83 (1). Wiley Online Library: 363–385.
29

**Supplementary Materials**

Table S1. Complete prompt created based on the geo-knowledge of common forms of location descriptions.

*This is a set of location description recognition problems.*
*The `Sentence` is a sentence containing location descriptions.*
*The goal is to infer which parts of the sentence represent location descriptions and the categories of the location descriptions. Split different location descriptions with `;`.*
*--*

*--*
*Sentence: Papa stranded in home. Water rising above waist. HELP 812 Wood Ln, 77828 #houstonflood*
*Q: Which parts of this sentence represent location descriptions?*
*A: C1: 812 Wood Ln, 77828*
*--*

*--*
*Sentence: Anyone doing high water rescues in the Pasadena/Deer Park area? My daughter has been stranded in a parking lot all night*
*Q: Which parts of this sentence represent location descriptions?*
*A: C10: Pasadena/Deer Park*
*--*

*--*
*Sentence: Allen Parkway, Memorial, Waugh overpass, Spotts park and Buffalo Bayou park completely under water*
*Q: Which parts of this sentence represent location descriptions?*
*A: C2: Allen Parkway; C2: Memorial; C2: Waugh overpass; C7: Spotts park; C7: Buffalo Bayou park*
*--*

*--*
*Sentence: Streets Flooded: Almeda Genoa Rd. from Windmill Lakes Blvd. to Rowlett Rd. HurricaneHarvey Houston*
*Q: Which parts of this sentence represent location descriptions?*
*A: C11: Almeda Genoa Rd. from Windmill Lakes Blvd. to Rowlett Rd.; C9: Houston*
*--*

*--*
*Sentence: Cleaning supply drive is underway. 9-11 am today at Preston Hollow Presbyterian Church*
*Q: Which parts of this sentence represent location descriptions?*
*A: C8: Preston Hollow Presbyterian Church*
*--*

*--*
*Sentence: #Harvey LIVE from San Antonio, TX. Fatal car accident at Ingram Rd., Strong winds.*

*Q: Which parts of this sentence represent location descriptions?*
*A: C9: San Antonio; C9: TX; C2: Ingram Rd.*
*--*

*--*
*Sentence: 9:00AM update video from Hogan St over White Oak Bayou, I-10, I-45: water down about 4' since last night.*
*Q: Which parts of this sentence represent location descriptions?*
*A: C5: Hogan St over White Oak Bayou; C3: I-10; C3: I-45*
*--*

*--*
*Sentence: Left Corpus bout to be in San Angelo #HurricaneHarvey Y'all be safe Avoided highway 37 Took the back road*
*Q: Which parts of this sentence represent location descriptions?*
*A: C9: Corpus; C9: San Angelo; C3: highway 37*
*--*

*--*
*Sentence: Need trailers/trucks to move dogs from Park Location: Whites Park Pavillion off I-10 exit 61 Anahuac TX*
*Q: Which parts of this sentence represent location descriptions?*
*A: C7: Whites Park Pavillion; C3: I-10; C4: exit 61; C9: Anahuac; C9: TX*
*--*

*--*
*Sentence: Townsend exit, Sorters road and Hamblen road is flooded coming from 59 southbound HurricaneHarvery Harvey2017*
*Q: Which parts of this sentence represent location descriptions?*
*A: C4: Townsend exit; C5: Sorters road and Hamblen road; C3: 59 southbound*
*--*

*--*
*Sentence: #Harvey does anyone know about the flooding conditions around Cypress Ridge High School?! #HurricaneHarvey*
*Q: Which parts of this sentence represent location descriptions?*
*A: C8: Cypress Ridge High School*
*--*

*--*
*Sentence: FYI to any of you in NW Houston/Lakewood Forest, Projections are showing Cypress Creek overflowing at Grant Rd*
*Q: Which parts of this sentence represent location descriptions?*

*A: C10: NW Houston/Lakewood Forest; C5: Cypress Creek overflowing at Grant Rd*
--

--
*Sentence: #HurricaneHarvey family needs rescuing at 11800 Grant Rd. Apt. 1009 in Cypress, Texas 77429, 2 elderly, one is 90 not steady in her feet*
*Q: Which parts of this sentence represent location descriptions?*
*A: C1: 11800 Grant Rd. Apt. 1009 in Cypress, Texas 77429*
--

--
*Sentence: Guys, this is I-45 @ Main Street in Houston. Crazy. #hurricane #harvey. . .*
*Q: Which parts of this sentence represent location descriptions?*
*A: C5: I-45 @ Main Street; C9: Houston*
--

--
*Sentence: Frontage Rd at the river #hurricaneHarvey #hurricaneharvey @ San Jacinto River*
*Q: Which parts of this sentence represent location descriptions?*
*A: C2: Frontage Rd; C6: San Jacinto River*
--

--
*Sentence: Pictures of downed trees and damaged apartment building on Airline Road in Corpus Christi.*
*Q: Which parts of this sentence represent location descriptions?*
*A: C2: Airline Road; C9: Corpus Christi*
--

--
*Sentence: Buffalo Bayou holding steady at 10,000 cfs at the gage near Terry Hershey Park*
*Q: Which parts of this sentence represent location descriptions?*
*A: C6: Buffalo Bayou; C7: Terry Hershey Park*
--

--
*Sentence: Major flooding at Clay Rd & Queenston in west Houston. Lots of rescues going on for ppl trapped...*
*Q: Which parts of this sentence represent location descriptions?*
*A: C5: Clay Rd & Queenston; C9: Houston*
--

--

*Sentence: HELP! A pregnant lady is stuck in her car on I-45 between Cypress Hill & Huffmeister exits #harvey*
*Q: Which parts of this sentence represent location descriptions?*
*A: C11: I-45 between Cypress Hill & Huffmeister exits*
*--*

*--*
*Sentence: If you need a place to escape #HurricaneHarvey, The Willie De Leon Civic Center: 300 E. Main St in Uvalde is open as a shelter*
*Q: Which parts of this sentence represent location descriptions?*
*A: C7: The Willie De Leon Civic Center; C1: 300 E. Main St in Uvalde*
*--*

*--*
*Sentence: Houston's Buffalo Bayou Park - always among the first to flood. #Harvey*
*Q: Which parts of this sentence represent location descriptions?*
*A: C9: Houston; C7: Buffalo Bayou Park*
*--*

*--*
*Sentence: #HurricaneHarvey INTENSE eye wall of category 4 Hurricane Harvey from Rockport, TX*
*Q: Which parts of this sentence represent location descriptions?*
*A: C9: Rockport; C9: TX*
*--*

*--*
*Sentence: {TEXT}*
*Q: Which parts of this sentence represent location descriptions?*
*A:*

Table S2. Complete prompt created based on a different set of 22 tweet examples using the same geo-knowledge.

*This is a set of location description recognition problems.*
*The `Sentence` is a sentence containing location descriptions.*
*The goal is to infer which parts of the sentence represent location descriptions and the categories of the location descriptions. Split different location descriptions with `;`.*
*--*

*--*
*Sentence: Pls rescue 12 day baby family at 7 Woodview St, Houston, 77124, flooding will reach roof soon…*

*Q: Which parts of this sentence represent location descriptions?*
*A: C1: 7 Woodview St, Houston, 77124*
*--*

*--*
*Sentence: What's left of my front and back yard in the League City/Dickinson area.*
*Q: Which parts of this sentence represent location descriptions?*
*A: C10: League City/Dickinson*
*--*

*--*
*Sentence: Thanks Home Depot in Plano on Custer Rd for your help with Flood Cleanup Kits!*
*Q: Which parts of this sentence represent location descriptions?*
*A: C9: Plano; C2: Custer Rd*
*--*

*--*
*Sentence: A short video of the aftermath of HurricaneHarvey in Port Aransas on Cotter Ave. from Alister to Station St., 9/3/...*
*Q: Which parts of this sentence represent location descriptions?*
*A: C9: Port Aransas; C11: Cotter Ave. from Alister to Station St.*
*--*

*--*
*Sentence: First St Baptist Church needs water & cleaning supplies for Port Arthur community flooded by Harvey*
*Q: Which parts of this sentence represent location descriptions?*
*A: C8: First St Baptist Church; C9: Port Arthur*
*--*

*--*
*Sentence: Can anyone get food to memorial Hermann hospital at Gessner Rd in Houston? Request from CajunNavy hurricaneHarvey ...*
*Q: Which parts of this sentence represent location descriptions?*
*A: C8: memorial Hermann hospital; C2: Gessner Rd; C9: Houston*
*--*

*--*
*Sentence: Garth road & IH10 at Baytown, Tx. Very High water. Baytown REPORT ON harvey2017*
*Q: Which parts of this sentence represent location descriptions?*
*A: C5: Garth road & IH10; C9: Baytown; C9: Tx; C9: Baytown*
*--*

--
*Sentence: someone placed a car here for ins purpose. parking lot of I-10 highway. Terry Hershey Park Parking lot. ...*
*Q: Which parts of this sentence represent location descriptions?*
*A: C3: I-10 highway; C7: Terry Hershey Park*
--

--
*Sentence: Need trailers/trucks to move dogs from Park Location: Whites Park Pavillion off I-10 exit 61 Anahuac TX*
*Q: Which parts of this sentence represent location descriptions?*
*A: C7: Whites Park Pavillion; C3: I-10; C4: exit 61; C9: Anahuac; C9: TX*
--

--
*Sentence: 59 north, towards the Polk St exit: Go to Convention Center. houstonflood KHOU11 288 texasflood*
*Q: Which parts of this sentence represent location descriptions?*
*A: C3: 59 north; C4: Polk St exit; C7: Convention Center*
--

--
*Sentence: Community is responding at shelters in College Park High School and Magnolia High School in TheWoodlands Harvey ...*
*Q: Which parts of this sentence represent location descriptions?*
*A: C8: College Park High School; C8: Magnolia High School; C9: TheWoodlands'*
--

--
*Sentence: FYI to any of you in NW Houston/Lakewood Forest, Projections are showing Cypress Creek overflowing at Grant Rd*
*Q: Which parts of this sentence represent location descriptions?*
*A: C10: NW Houston/Lakewood Forest; C5: Cypress Creek overflowing at Grant Rd*
--

--
*Sentence: Father in law in a wheelchair. Send help to 6312 Bapling Drive, Sugar Land, TX HarveyStorm HarveyRelief HARVEYHELP HoustonStrande*
*Q: Which parts of this sentence represent location descriptions?*
*A: C1: 6312 Bapling Drive, Sugar Land, TX*
--

--

*Sentence: Gas is apparently going up this week in San Antonio because of HurricaneHarvey, it's still 2.51 at the co-op on 17th & 36th ave*
*Q: Which parts of this sentence represent location descriptions?*
*A: C9: San Antonio; C5: 17th & 36th ave*
--

--
*Sentence: White Oak Bayou around Stude Park. HoustonFloods Houston Heights*
*Q: Which parts of this sentence represent location descriptions?*
*A: C6: White Oak Bayou; C7: Stude Park; C9: Houston*
--

--
*Sentence: Worried. A foot 2 go before Oyster Creek starts spilling into Lakes of Brightwater, Lakefront Drive, in front of our house ...*
*Q: Which parts of this sentence represent location descriptions?*
*A: C6: Oyster Creek; C6: Lakes of Brightwater; C2: Lakefront Drive*
--

--
*Sentence: Texas City asking 18 homes on S. Humble Camp Rd. to evacuate in case GCWA reservoir breaches Harvey*
*Q: Which parts of this sentence represent location descriptions?*
*A: C9: Texas City; C2: S. Humble Camp Rd.; C7: GCWA reservoir*
--

--
*Sentence: Major flooding at Clay Rd & Queenston in west Houston. Lots of rescues going on for ppl trapped...*
*Q: Which parts of this sentence represent location descriptions?*
*A: C5: Clay Rd & Queenston; C9: Houston*
--

--
*Sentence: Closed due to train derailment in Harvey on Destrehan Ave between River Rd and 4th St traffic NOLA*
*Q: Which parts of this sentence represent location descriptions?*
*A: C11: Destrehan Ave between River Rd and 4th St*
--

--
*Sentence: Drop Off Location: The Life Center, Charlotte, 2435 Toomey Ave. #SocksForHouston LCFellowship ...*

*Q: Which parts of this sentence represent location descriptions?*
*A: C7: The Life Center; C9: Charlotte; C1: 2435 Toomey Ave*
*--*

*--*
*Sentence: Do we know how Minute Maid Park and NRG Park are?? Are they flooded as well as Honda Center? hurricaneharvey*
*Q: Which parts of this sentence represent location descriptions?*
*A: C7: Minute Maid Park; C7: NRG Park; C7: Honda Center*
*--*

*--*
*Sentence: Eye wall still very much intact as Harvey is nearly stationary north of Rockport, TX. It's going to be a long nigh ...*
*Q: Which parts of this sentence represent location descriptions?*
*A: C9: Rockport; C9: TX*
*--*

*--*
*Sentence: {TEXT}*
*Q: Which parts of this sentence represent location descriptions?*
*A:*

Table S3. Complete prompt created based on a set of 22 synthesized tweet examples using the same geo-knowledge.

*This is a set of location description recognition problems.*
*The `Sentence` is a sentence containing location descriptions.*
*The goal is to infer which parts of the sentence represent location descriptions and the categories of the location descriptions. Split different location descriptions with `;`.*
*--*

*--*
*Sentence: Please help family needs rescue at 112 Wikleson Dr, Houston 42143*
*Q: Which parts of this sentence represent location descriptions?*
*A: C1: 112 Wikleson Dr, Houston 42143*
*--*

*--*
*Sentence: Heavy flooding in the Diego City/Amherst area. Please don't go there*
*Q: Which parts of this sentence represent location descriptions?*
*A: C10: Diego City/Amherst*

--

--
*Sentence: Home Depot on Niagara Rd is donating tools for house repair #Harvey*
*Q: Which parts of this sentence represent location descriptions?*
*A: C2: Niagara Rd*
--

--
*Sentence: There is an accident in Port Aransas on Bailey Rd. between Airport Dr and Station St*
*Q: Which parts of this sentence represent location descriptions?*
*A: C9: Port Aransas; C11: Bailey Rd. between Airport Dr and Station St*
--

--
*Sentence: Water & cleaning supplies are needed at Grace Family Bible Church. Please help out...*
*Q: Which parts of this sentence represent location descriptions?*
*A: C8: Grace Family Bible Church*
--

--
*Sentence: Help needed at memorial Hermann hospital at Gessner Rd in Houston! People need water and food there hurricaneHarvey ...*
*Q: Which parts of this sentence represent location descriptions?*
*A: C8: memorial Hermann hospital; C2: Gessner Rd; C9: Houston*
--

--
*Sentence: The intersection of Main road & I-60 is flooded. Very High water. harvey2017*
*Q: Which parts of this sentence represent location descriptions?*
*A: C5: Main road & I-60*
--

--
*Sentence: A person is trapped at the parking lot off I-10 highway near Buffalo Bayou park ...*
*Q: Which parts of this sentence represent location descriptions?*
*A: C3: I-10 highway; C7: Buffalo Bayou park*
--

--
*Sentence: Exit 53 and exit 54 of I-11 are both flooded. Please avoid these two exits. Houston*
*Q: Which parts of this sentence represent location descriptions?*
*A: C4: Exit 53; C4: exit 54; C3: I-11; C9: Houston*

--

--
*Sentence: 63 south, towards the Main St exit: Go to Convention Center. houstonflood 288 texasflood*
*Q: Which parts of this sentence represent location descriptions?*
*A: C3: 63 south; C4: Main St exit; C7: Convention Center*
--

--
*Sentence: Shelters are provided in Williamsville High School. Please go there if you need a place to stay ...*
*Q: Which parts of this sentence represent location descriptions?*
*A: C8: Williamsville High School*
--

--
*Sentence: For those of you living in NW Houston/Lakewood Forest, Grant Rd @ Cypress Rd is flooded. Avoid it!*
*Q: Which parts of this sentence represent location descriptions?*
*A: C10: NW Houston/Lakewood Forest; C5: Grant Rd @ Cypress Rd*
--

--
*Sentence: Grandpa needs help at 1831 West Ridge Rd, Sugar Land, TX HarveyStorm HarveyRelief HARVEYHELP HoustonStrande*
*Q: Which parts of this sentence represent location descriptions?*
*A: C1: 1831 West Ridge Rd, Sugar Land, TX*
--

--
*Sentence: High water at the 11th & 30th ave. Houston is battered by heavy rain*
*Q: Which parts of this sentence represent location descriptions?*
*A: C5: 11th & 30th ave; C9: Houston*
--

--
*Sentence: White Oak Bayou around Clear Lake Park. HoustonFloods Houston Heights*
*Q: Which parts of this sentence represent location descriptions?*
*A: C6: White Oak Bayou; C7: Clear Lake Park; C9: Houston*
--

--
*Sentence: This is Buffalo Creek running under Kingsview Bridge near Frontier Drive, crazy ...*

*Q: Which parts of this sentence represent location descriptions?*
*A: C6: Buffalo Creek; C7: Kingsview Bridge; C2: Frontier Drive*
--

--
*Sentence: Families living on S. Washington Rd. will have to evacuate before the water gets too high #Harvey*
*Q: Which parts of this sentence represent location descriptions?*
*A: C2: S. Washington Rd.*
--

--
*Sentence: Major flooding at High Rd & 15 Ave. in west Houston. People trapped. Please help...*
*Q: Which parts of this sentence represent location descriptions?*
*A: C5: High Rd & 15 Ave.; C9: Houston*
--

--
*Sentence: Highway 10 between River Rd and 4th St is flooded. Please use local roads NOLA*
*Q: Which parts of this sentence represent location descriptions?*
*A: C11: Highway 10 between River Rd and 4th St*
--

--
*Sentence: If you have things to donate, you can drop them off at ClearField Community Center, Houston, 351 7th Ave.*
*Q: Which parts of this sentence represent location descriptions?*
*A: C7: ClearField Community Center; C9: Houston; C1: 351 7th Ave*
--

--
*Sentence: Hurricane eye wall is approaching Port Aransas, TX. Pray...*
*Q: Which parts of this sentence represent location descriptions?*
*A: C9: Port Aransas; C9: TX*
--

--
*Sentence: Do we know the flooding conditions at Buffalo Bayou Park and Minute Maid Park? Any information can help.*
*Q: Which parts of this sentence represent location descriptions?*
*A: C7: Buffalo Bayou Park; C7: Minute Maid Park*
--

> --
> *Sentence: {TEXT}*
> *Q: Which parts of this sentence represent location descriptions?*
> *A:*

Table S4. Results of the Geo-GPT-4 model based on two additional sets of tweet examples.

| **Models** | **Precision** | **Recall** | **F-score** |
|---|---|---|---|
| *Geo-GPT-4_set2\** | 0.701 | 0.737 | 0.718 |
| *Geo-GPT-4_syn* | 0.699 | 0.707 | 0.703 |

\* Note: Because a different set of 22 tweets are used in the prompt for this model, the precision, recall, and F-score of *Geo-GPT-4_set2* are calculated based on a slightly different set of 978 tweets excluding those 22 tweets used in the prompt.

|  | C1 | C2 | C3 | C4 | C5 | C6 | C7 | C8 | C9 | C10 | C11 | Missed |
|---|---|---|---|---|---|---|---|---|---|---|---|---|
| Door numbers: C1 | 240<br>91.25% | 1<br>0.38% | 0<br>0.00% | 0<br>0.00% | 0<br>0.00% | 0<br>0.00% | 0<br>0.00% | 1<br>0.38% | 2<br>0.76% | 0<br>0.00% | 0<br>0.00% | 19<br>7.22% |
| Street names: C2 | 0<br>0.00% | 145<br>64.73% | 0<br>0.00% | 0<br>0.00% | 7<br>3.12% | 0<br>0.00% | 1<br>0.45% | 0<br>0.00% | 1<br>0.45% | 0<br>0.00% | 0<br>0.00% | 70<br>31.25% |
| Highways: C3 | 0<br>0.00% | 0<br>0.00% | 13<br>50.00% | 0<br>0.00% | 1<br>3.85% | 0<br>0.00% | 0<br>0.00% | 0<br>0.00% | 0<br>0.00% | 0<br>0.00% | 0<br>0.00% | 12<br>46.15% |
| Exits of highways: C4 | 0<br>0.00% | 0<br>0.00% | 0<br>0.00% | 2<br>33.33% | 0<br>0.00% | 0<br>0.00% | 0<br>0.00% | 0<br>0.00% | 0<br>0.00% | 0<br>0.00% | 0<br>0.00% | 4<br>66.67% |
| Intersections: C5 | 0<br>0.00% | 8<br>7.69% | 1<br>0.96% | 0<br>0.00% | 71<br>68.27% | 3<br>2.88% | 0<br>0.00% | 0<br>0.00% | 0<br>0.00% | 1<br>0.96% | 1<br>0.96% | 19<br>18.27% |
| Natural features: C6 | 0<br>0.00% | 0<br>0.00% | 0<br>0.00% | 0<br>0.00% | 0<br>0.00% | 39<br>63.93% | 1<br>1.64% | 0<br>0.00% | 3<br>4.92% | 12<br>19.67% | 0<br>0.00% | 6<br>9.84% |
| Other human-made: C7 | 1<br>0.45% | 12<br>5.38% | 0<br>0.00% | 0<br>0.00% | 0<br>0.00% | 3<br>1.35% | 143<br>64.13% | 10<br>4.48% | 4<br>1.79% | 21<br>9.42% | 0<br>0.00% | 29<br>13.00% |
| Local organizations: C8 | 0<br>0.00% | 0<br>0.00% | 0<br>0.00% | 0<br>0.00% | 0<br>0.00% | 0<br>0.00% | 1<br>1.59% | 53<br>84.13% | 0<br>0.00% | 1<br>1.59% | 0<br>0.00% | 8<br>12.70% |
| Admin units: C9 | 0<br>0.00% | 1<br>0.14% | 0<br>0.00% | 0<br>0.00% | 3<br>0.42% | 0<br>0.00% | 2<br>0.28% | 4<br>0.56% | 463<br>65.21% | 57<br>8.03% | 0<br>0.00% | 180<br>25.35% |
| Multiple areas: C10 | 0<br>0.00% | 0<br>0.00% | 0<br>0.00% | 0<br>0.00% | 0<br>0.00% | 0<br>0.00% | 0<br>0.00% | 0<br>0.00% | 0<br>0.00% | 5<br>100.00% | 0<br>0.00% | 0<br>0.00% |
| Road segments: C11 | 0<br>0.00% | 2<br>22.22% | 0<br>0.00% | 0<br>0.00% | 3<br>33.33% | 0<br>0.00% | 0<br>0.00% | 0<br>0.00% | 0<br>0.00% | 0<br>0.00% | 1<br>11.11% | 3<br>33.33% |
| Not in ground truth | 26<br>6.70% | 32<br>8.25% | 9<br>2.32% | 2<br>0.52% | 38<br>9.79% | 12<br>3.09% | 47<br>12.11% | 40<br>10.31% | 123<br>31.70% | 44<br>11.34% | 4<br>1.03% | 11<br>2.84% |

Figure S1. Confusion matrix based on the output of Geo-GPT-4 and the annotated location descriptions using relaxed matching with 75% overlapping.